\documentclass[a4paper,11pt]{article}
\pdfoutput=1 

\usepackage{jheppub} 
\usepackage{hyperref}
\usepackage{caption}
\usepackage{subcaption}
\usepackage[T1]{fontenc} 
\usepackage{listings}

\title{Higher-loop integrated negative geometries in ABJM}
\author[a,b]{Mart\'in Lagares}
\author[b]{Shun-Qing Zhang}

\affiliation[a]{Instituto de Fisica La Plata, Universidad Nacional de La Plata,
C.C. 67, 1900 La Plata, Argentina}
\affiliation[b]{Max-Planck-Institut f\"{u}r Physik, Werner-Heisenberg-Institut, D-80805 M\"{u}nchen, Germany}

\emailAdd{martinlagares95@gmail.com, sqzhang@mpp.mpg.de}

\preprint{MPP-2024-37}

\abstract{
In the three-dimensional ${\cal N}=6$ Chern-Simons matter (ABJM) theory, the integrand for the logarithm of the scattering amplitude admits a decomposition in terms of negative geometries, which implies that all the infrared divergences concentrate in the last loop integration. 
We compute the infrared-finite functions that arise from performing a three-loop integration over the four-loop integrand for the logarithm of the four-point amplitude, for which we use the method of differential equations. Our results provide a direct computation of the four-loop cusp anomalous dimension of the theory, in agreement with the current all-loop integrability-based proposal. We find an apparent simplicity in the leading singularities of the integrated results, provided one works in the frame in which the unintegrated loop variable goes to infinity. Finally, our results suggest an alternating sign pattern for the integrated negative geometries in the Euclidean region.

}

\begin{document} 
\maketitle

\section{Introduction}

The study of scattering amplitudes has unveiled remarkable structures that are satisfied by these observables, which are obscured in the standard perturbative expansion in terms of Feynman diagrams. One striking step in this direction was made in \cite{Arkani-Hamed:2013jha}, where a geometric description of scattering amplitudes was proposed in the context of the planar ${\cal N}=4$ super Yang Mills (sYM) theory. More precisely, it was shown that the $L$-loop integrand for the $n$-point scattering amplitude in the planar limit of ${\cal N}=4$ sYM can be obtained as the canonical form of a certain \textit{positive geometry}, the Amplituhedron
\cite{Arkani-Hamed:2013jha,Arkani-Hamed:2013kca,Franco:2014csa,Arkani-Hamed:2017vfh,Arkani-Hamed:2018rsk,Damgaard:2019ztj,Ferro:2022abq,Ferro:2020ygk,Herrmann:2022nkh}. This result provided an alternative perturbative framework for the study of scattering amplitudes, and was later generalized to various contexts, including non-supersymmetric theories \cite{Arkani-Hamed:2017mur} and cosmology \cite{Arkani-Hamed:2017fdk}. In particular, a similar geometric description was recently proposed \cite{Huang:2021jlh,He:2021llb,He:2022cup,He:2023rou,Lukowski:2023nnf} for the three-dimensional ${\cal N}=6$ super Chern-Simons theory known as ABJM \cite{Aharony:2008ug}.

Performing loop integrals is usually a bottleneck step in the computation of scattering amplitudes, and when doing so one often has to deal with infrared (IR) divergences\footnote{Recently, a geometry-based approach to regulate IR divergences was initiated in \cite{Arkani-Hamed:2023epq}.}. In \cite{Arkani-Hamed:2021iya,He:2022cup,He:2023rou} it was proposed that, both for the ${\cal N}=4$ sYM and the ABJM theories, the integrand for the logarithm of the scattering amplitude is described by the canonical forms of \textit{negative} geometries. Interestingly, this description in terms of negative geometries was shown to imply that the result of performing $L-1$ of the loop integrals over the $L$-loop integrand for the logarithm of the amplitude is IR-finite \cite{Arkani-Hamed:2021iya}, i.e. all the IR divergences concentrate on the last loop integral. In other words, the geometric description provides an IR-finite result that is one integral away from the logarithm of the amplitude, and therefore motivates its study.

The result of performing $L-1$ of the loop integrations over the $L$-loop integrand for the logarithm of the amplitude has been previously studied at four points both in ${\cal N}=4$ sYM \cite{Arkani-Hamed:2021iya,Alday:2011ga,Adamo:2011cd,Engelund:2011fg,Alday:2012hy,Alday:2013ip,Henn:2019swt,Chicherin:2022bov} and in ABJM \cite{Henn:2023pkc,He:2023exb}, up to $L=4$ and $L=3$ respectively. Based on the Wilson loops/scattering amplitudes duality \cite{Alday:2007hr,Drummond:2007aua,Brandhuber:2007yx,Drummond:2007cf,Henn:2010ps,Bianchi:2011dg,Chen:2011vv,Leoni:2015zxa}, prescriptions have been proposed to compute the cusp anomalous dimension $\Gamma_{\rm cusp}$ of each theory by applying certain functionals on the integrated results \cite{Arkani-Hamed:2021iya,Alday:2013ip,Henn:2019swt,Henn:2023pkc,He:2023exb}. This has allowed to obtain the full four-loop value of $\Gamma_{\rm cusp}$ in ${\cal N}=4$ sYM, including the first non-planar corrections \cite{Henn:2019swt}. In the ABJM case, these prescriptions were used to compute the three-loop contribution to $\Gamma_{\rm cusp}$ \cite{Henn:2023pkc,He:2023exb}, in agreement with the all-loop integrability-based proposals \cite{Gromov:2008qe}. Moreover, the integrated results have been shown to have uniform transcendental weight in their coupling expansion \cite{Arkani-Hamed:2021iya,Alday:2011ga,Alday:2012hy,Alday:2013ip,Henn:2019swt,Chicherin:2022bov,Henn:2023pkc,He:2023exb}, and the corresponding leading singularities have been proven to be conformally invariant \cite{Chicherin:2022bov,Henn:2023pkc}. Remarkably, all-loop sums have been obtained for certain subsets of integrated negative geometries in the ${\cal N}=4$ sYM theory \cite{Arkani-Hamed:2021iya} (more specifically, for the ladder and tree diagrams in the diagrammatic notation of \cite{Arkani-Hamed:2021iya}). Higher-point integrated results have also been studied for the ${\cal N}=4$ sYM case, up to $L=3$ for five particles \cite{Chicherin:2022zxo} and up to $L=2$ for arbitrary number of particles \cite{Chicherin:2022bov}. Interestingly, this analysis has suggested a duality between the integrated results in ${\cal N}=4$ sYM and all-plus scattering amplitudes in the pure Yang-Mills theory \cite{Chicherin:2022bov}. Finally, strong-coupling results have been obtained in the ${\cal N}=4$ sYM case for the four- and six-particle cases \cite{Alday:2011ga,Hernandez:2013kb}.

The above discussion poses the natural problem of computing the integrated negative geometries to higher loops. In this paper we give a step forward in that direction by computing the $L=4$ contribution to the four-point integrated negative geometries of the ABJM theory. A better understanding of the higher-loop structure of the integrated negative geometries of ABJM could shed light on an all-loop computation of the cusp anomalous dimension $\Gamma_{\rm cusp}$ of the theory\footnote{A Thermodynamic Bethe Ansatz approach for the computation of $\Gamma_{\rm cusp}$ was recently studied for the ABJM theory in \cite{Correa:2023lsm}.}. In turn, this knowledge of $\Gamma_{\rm cusp}$ could give insight on the non-perturbative structure of the interpolating function $h(\lambda)$ of ABJM \cite{Beisert:2006qh,Minahan:2008hf,Minahan:2009te,Bak:2008vd,Minahan:2009aq,Minahan:2009wg,Leoni:2010tb,Gromov:2014eha,Cavaglia:2016ide}, which percolates in every integrability-based computation made in this three-dimensional theory. An all-loop proposal for $h(\lambda)$ was made in \cite{Gromov:2014eha}.

Following the diagrammatic representation of \cite{Arkani-Hamed:2021iya}, negative geometries organize scattering amplitudes of the ABJM theory in a perturbative expansion in terms of \textit{connected} and \textit{bipartite} graphs \cite{He:2022cup,He:2023rou}. It is at the $L=4$ loop order that one encounters the first ``loops of loops'' diagram in the expansion of the four-point negative geometry \cite{He:2022cup} \footnote{For the negative geometries in the ${\cal N}=4$ sYM theory the first ``loops of loops'' diagram shows up at $L=3$ \cite{Arkani-Hamed:2021iya,Brown:2023mqi}. The canonical forms of all one-cycle geometries are given in \cite{Brown:2023mqi}.}. More precisely, at this perturbative order one has to consider a box diagram, which contributes with a non-trivial three-loop integral to our computation. To perform this and other loop integrals we have used the method of differential equations \cite{Henn:2013pwa}, that has been widely applied to compute loop integrals in perturbative quantum field theory. 

As we will show, the integrated results have uniform transcendentality up to $L=4$ loops, with transcendental weight $L-1$ for odd values of $L$ and with weight $L-2$ for even values of $L$. Moreover, at the integrated level the results follow an alternating sign pattern within the Euclidean region. Remarkably, the leading singularities are restricted to the set $\{ s\sqrt{t}, \, t\sqrt{s}, \, \sqrt{st(s+t)} \}$ in the limit at which the unintegrated loop variable goes to infinity.
Finally, we will use the $L=4$ integrated negative geometry to compute the four-loop contribution to the cusp anomalous dimension of the theory. This result is the first explicit four-loop computation of $\Gamma_{\rm cusp}$ in the ABJM theory, and it agrees with the all-loop integrability-based proposal \cite{Gromov:2008qe}.

The plan of the paper is as follows. In Section \ref{sec review} we will give a short review of the Amplituhedron construction in the ABJM theory, with a focus on the current results for the corresponding integrated negative geometries. In Section \ref{sec three loop results} we will present the computation of the $L=4$ integrated negative geometry, and in Section \ref{sec properties} we will discuss the main properties of this result. Section \ref{sec gamma cusp} is devoted to the computation of the four-loop contribution to the cusp anomalous dimension of the theory. Finally, we give our conclusions in Section \ref{sec conclusion}. We also include an Appendix and ancillary files that complement the results presented in the main body of the paper.

\section{Geometric description of amplitudes in the ABJM theory}
\label{sec review}

In the reference \cite{Arkani-Hamed:2013jha}, a description of amplitudes in terms of positive geometries was proposed for the ${\cal N}=4$ sYM theory. In turn, it was later shown that this result provided a description of the logarithm of the amplitude in terms of negative geometries \cite{Arkani-Hamed:2021iya}, which allowed to prove that the result of performing $L-1$ loop integrations over the $L$ loop integrand for the logarithm of the amplitude is free of IR divergences. The geometric construction of amplitudes has also been extended to the ABJM theory \cite{Huang:2021jlh,He:2021llb, He:2022cup,He:2023rou,Lukowski:2023nnf}. Moreover, the analysis of integrated negative geometries in ABJM was performed up to $L=3$ loops in \cite{Henn:2023pkc,He:2023exb}. This section is devoted to a short review of the ABJM Amplituhedron and its corresponding integrated negative geometries.

\subsection{The ABJM Amplituhedron}

Throughout this paper we will focus on a four-particle scattering process, with external momenta $p_i, \, i=1, \, \dots, \, 4$. To describe the kinematic data we will usually recur to dual-space coordinates ($x_i=p_{i+1}-p_i , \, i=1, \, \dots, \, 4$), for which we will often use five-dimensional notation.  We will employ capital $X$ letters when using five-dimensional notation, and we refer to the Appendix A of \cite{Henn:2023pkc} for a discussion on its properties. Alternatively, we will also use momentum-twistors \cite{Hodges:2009hk} to describe the kinematics.

Let us start by defining the $L$-loop integrands ${\cal I}_L$ and ${\cal L}_L$ for the amplitude and its logarithm as 
\begin{equation}
\label{amplitude integrand}
{\cal M} \bigg|_{L {\rm  \, loops}} := \left( \prod_{j=5}^{4+L} \int \frac{d^3 x_j}{i\pi^{3/2}} \right) \, {\cal I}_L \,, \qquad 
\log {\cal M} \bigg|_{L{\rm  \, loops}} := \left( \prod_{j=5}^{4+L} \int \frac{d^3 x_j}{i\pi^{3/2}} \right) \, {\cal L}_L \,,
\end{equation}
where $x_5,x_6, \dots, x_{4+L}$ describe the loop variables and ${\cal M} := {\cal A}/{\cal A}_{\rm tree}$, with ${\cal A}$ the color-ordered scattering amplitude and ${\cal A}_{\rm tree}$ its corresponding tree-level value. In \cite{He:2022cup} it was proposed that the $L$-loop integrand ${\cal I}_L$ for the four-particle ABJM amplitude can be obtained as
\begin{equation}
{\cal I}_L= n_L \, \Omega_L \,,
\end{equation}
where $\Omega_L$ is the canonical form of a positive geometry known as the $L$-loop ABJM Amplituhedron, and $n_L$ is a normalization given as\footnote{We are using the convention $\lambda = N/k$
for the 't Hooft coupling of the theory, with $N$ the number of colors and $k$ the Chern-Simons level.}
\begin{equation}
    \label{relative normalizations}
    n_L = \frac{1}{L!} \left( \frac{i}{2\sqrt{\pi}}\right)^L \,.
\end{equation}
We refer to Appendix B of \cite{Henn:2023pkc} for a discussion on the above relative normalization. In order to define the $L$-loop ABJM Amplituhedron we need to consider the region in momentum-twistor space characterized by the constraints
\begin{align}
\label{Amplituhedron constr 1}
    &\langle 1234 \rangle <0 \,, \\
    \label{Amplituhedron constr 2}
    \langle l_i 12 \rangle <0 ,\, \quad \langle l_i 23 \rangle &<0 ,\, \quad \langle l_i 34 \rangle <0 ,\, \quad \langle l_i 14 \rangle <0 \,, \\
    \label{Amplituhedron constr 3}
    \langle l_i 13 \rangle &>0 ,\, \quad \langle l_i 24 \rangle >0 \,, \\
    \label{Amplituhedron constr 4}
    & \, \, \langle l_i l_j \rangle <0 \,,
\end{align}
for all $i,j=5,\, \dots,\, 4+L$. Above we are using the notation $\langle klmn \rangle=\epsilon_{IJKL} Z^I_k Z^J_l Z^K_m Z^L_n$, where $Z_i, \, i=1, \, \dots , \, 4$ are the momentum-twistors that characterize the external kinematic data and $l_5^{IJ}:=Z_A^I Z_B^J$, $l_6^{IJ}:=Z_C^I Z_D^J$, $l_7^{IJ}:=Z_E^I Z_F^J$, $\dots$ are the lines in momentum-twistor space that describe the loop variables. Let us also consider the symplectic constraints given by
\begin{align}
\label{constraint tree level}
\Sigma_{IJ} Z_i^I Z_{i+1}^J &=  0 \,, \\
\label{L loop constraints}
\Sigma_{IJ} l^{IJ}_k &=0 \,,
\end{align}
for all $i=1,\, \dots,\, 4$ and $k=5,\, \dots,\, 4+L$ all, where $\Sigma$ is defined as
\begin{equation}
\label{Sigma}
\Sigma= \left( 
\begin{array}{cc} 
0 & \epsilon_{2 \times 2} \\
\epsilon_{2 \times 2} & 0
\end{array}
 \right) \,,
\end{equation}
with $\epsilon_{2 \times 2}$ a totally anti-symmetric tensor. The $L$-loop ABJM Amplituhedron is then defined as the region in momentum-twistor space characterized by \eqref{Amplituhedron constr 1}-\eqref{L loop constraints}.

Interestingly, the above geometry imposes strong constraints on the $L$-loop integrand ${\cal L}_L$ for the logarithm of the amplitude. In order to discuss such constraints it is useful to consider the diagrammatic notation of \cite{Arkani-Hamed:2021iya}. We will use a node to represent a one-loop geometry defined by the constraints \eqref{Amplituhedron constr 1}-\eqref{Amplituhedron constr 3}, and we will recur to red thick lines to indicate a negativity condition between the loop momenta associated to two nodes, i.e.\footnote{Let us note that the negativity condition \eqref{mutual negativity} has the opposite sign to the one of the ${\cal N}=4$ sYM case \cite{Arkani-Hamed:2021iya}. In fact, all signs are reversed in the ABJM Amplituhedron with respect to the ${\cal N}=4$ sYM geometry. This is due to the fact that the symplectic condition \eqref{constraint tree level} implies $\langle 1234 \rangle <0$ for real-valued momentum-twistors \cite{He:2022cup,He:2023rou}.}
\begin{equation}
    \label{mutual negativity}
    \langle l_i l_j \rangle >0 \,,
\end{equation}
Then, the results of \cite{He:2022cup} imply that the $L$-loop integrand ${\cal L}_L$ for the logarithm of the amplitude is given as
\begin{equation}
\label{normalization L_L}
{\cal L}_L= \tilde{n}_L \, \tilde{\Omega}_L \,,
\end{equation}
with \cite{Henn:2023pkc}
\begin{equation}
\label{tilde n L}
\tilde{n}_L= \frac{1}{L!} \left( \frac{i}{2\sqrt{\pi}}\right)^L \,,
\end{equation}
and where the canonical form $\tilde{\Omega}_L$ can be obtained as 
\begin{equation}
\includegraphics[scale=0.3]{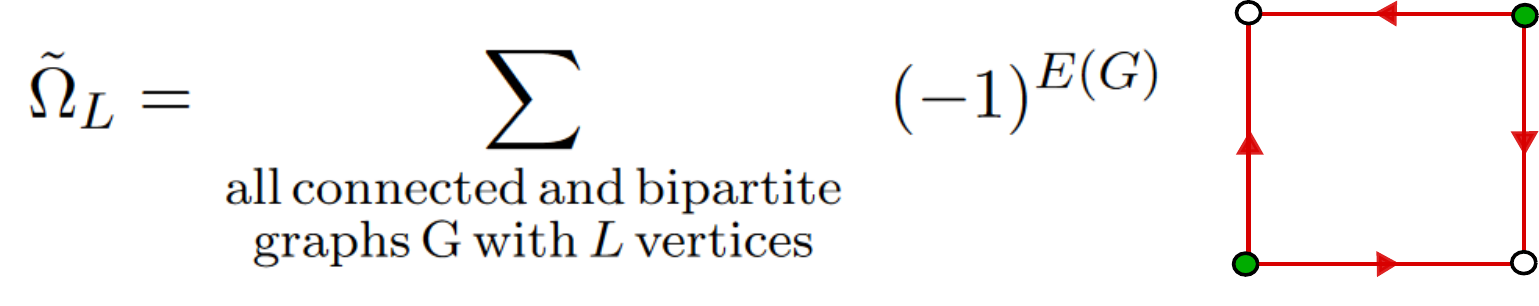} \label{bipartite expansion}
\end{equation}
with $E(G)$ the number of edges of each graph $G$ and where we are summing over the canonical forms of all possible \textit{connected} and \textit{bipartite} graphs that connect $L$ nodes with negativity conditions. We define a bipartite graph as one in which one can assign an orientation to each line in such a way that each node behaves either as a source (green nodes) or as a sink (white nodes). In comparison to the ${\cal N}=4$ sYM case, in which at $L$ loops one has to consider every possible connected graph with $L$ nodes, the restriction to bipartite graphs that appears in the ABJM case implies a significant reduction in the number of diagrams at each loop order. Using \eqref{bipartite expansion}, the canonical forms $\tilde{\Omega}_L$ were computed up to $L \leq 5$ in \cite{He:2022cup}.

Finally, let us note that the above construction was extended to all loops and all multiplicities in \cite{He:2023rou}. There the authors used the proposed ABJM Amplituhedron to obtain explicit results for the integrands for $n \leq 10$ and $n \leq 8$ particles at $L=1$ and $L=2$, respectively.

\subsection{Integrated negative geometries for $L \leq 3$}

As discussed in \cite{Arkani-Hamed:2021iya}, one advantage of expressing the $L$-loop integrand ${\cal L}_L$ in terms of negative geometries is that it implies that the result of integrating $L-1$ of the loop momenta is free of IR divergences. Following the ideas of \cite{Alday:2011ga}, it was shown in \cite{Henn:2023pkc,He:2023exb} that dual conformal symmetry\footnote{Evidence of dual-superconformal and Yangian symmetry has been found for ABJM amplitudes in \cite{Huang:2010qy,Gang:2010gy,Chen:2011vv,Bargheer:2010hn,Lee:2010du}.} constrains the integrated result to be
\begin{equation}
\label{n=4 general expression D=3}
\left( \prod_{j=6}^{4+L} \int \frac{d^3 X_j}{i\pi^{3/2}} \right) \, {\cal L}_L = \sqrt{\pi} \, \left( \frac{X_{13}^2 X_{24}^2}{X_{15}^2 X_{25}^2 X_{35}^2 X_{45}^2} \right)^{\frac{3}{4}} {\cal F}_{L-1}\left( z \right)  + \frac{i \, \epsilon\left( 1,2,3,4,5 \right)}{X_{15}^2 X_{25}^2 X_{35}^2 X_{45}^2} \, \frac{{\cal G}_{L-1} \left( z \right)}{\sqrt{\pi}} \,,
\end{equation}
with\footnote{Let us note that we are changing the definition of $z$ with respect to the conventions of \cite{Henn:2023pkc}. Moreover, we are also using a different normalization for the $\cal{F}$ function.}
\begin{equation}
\label{cross ratio}
z= \frac{X_{15}^2 X_{35}^2 X_{24}^2}{X_{25}^2 X_{45}^2 X_{13}^2} \,,
\end{equation}
and where
\begin{equation}
\epsilon(1,2,3,4,5):=\epsilon_{\mu \nu \rho \sigma \eta} X_1^{\mu} X_2^{\nu} X_3^{\rho} X_4^{\sigma} X_5^{\eta} \,.
\end{equation}
Dual conformal symmetry does not further constrain the functions ${\cal F}_{L-1}$ and ${\cal G}_{L-1}$, which should be computed by integration of the canonical forms $\tilde{\Omega}_L$. Those forms are given, up to $L \leq 3$, as \cite{He:2022cup}
\begin{equation}
    \label{amplituhedron canonical forms}
\includegraphics[scale=0.27]{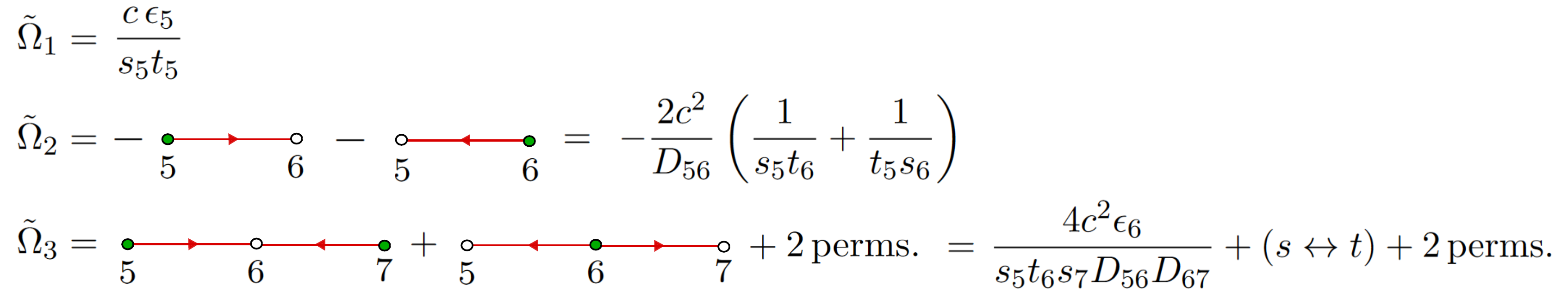}
\end{equation}
with
\begin{align}
\label{s,t,c,D,epsilon definitions}
s_i &:= \langle l_i 12 \rangle \langle l_i 34 \rangle \,,  \quad \quad
t_i := \langle l_i 23 \rangle \langle l_i 14 \rangle \,, \qquad \qquad \qquad
D_{ij} := -\langle l_i l_j \rangle \,, \\
c &:= \langle 1234 \rangle \,, \qquad \qquad
\epsilon_i := \sqrt{\langle l_i 13 \rangle \langle l_i 24 \rangle \langle 1234 \rangle} \,,
\end{align}
and where the permutations are over all nonequivalent configurations of the loop momenta. Let us note that up to $L \leq 3$ the only contributions to $\tilde{\Omega}_L$ come from \textit{ladder} diagrams, i.e. graphs in which all nodes are concatenated within a chain. By performing direct integration over \eqref{amplituhedron canonical forms} it was shown in \cite{Henn:2023pkc,He:2023exb} that
\begin{align}
\label{tree level F and G}
{\cal F}_0(z) &= 0\,, \; &{\cal G}_0(z) &= -2 \,, \\
\label{one loop F and G}
{\cal F}_1 (z) &= -\frac{1}{4} \left( z^{1/4} + \frac{1}{z^{1/4}} \right) \,, \; &{\cal G}_1 (z) &=0 \,, \\
\label{two loop F and G}
{\cal F}_2(z) &=0 \,, \; 
&{\cal G}_2(z)&=\frac{2}{3} \left[ \frac{{\cal H}(z) + \pi^2}{\sqrt{1+z}} \,  +  \frac{\pi^2}{2} + \left( z \to \frac{1}{z} \right)  \right] \,,
\end{align}
where
\begin{equation}
\begin{aligned}
{\cal H}(z) &=  -\text{Li}_2\left(\frac{2 \left(\sqrt{z+1}-1\right)}{z}\right)+\text{Li}_2\left(-\frac{2
   \left(\sqrt{z+1}+1\right)}{z}\right)\\
   & \quad \, +2 \log \left(\frac{4}{z}\right) \log
   \left(\frac{\sqrt{z+1}+1}{\sqrt{z}}\right)\,.
   \label{H-app}
\end{aligned}
\end{equation}

\section{$L=4$ integrated negative geometry}
\label{sec three loop results}

In this section we will turn to the main goal of this paper, the three-loop integration of the $L=4$ negative geometry of the ABJM theory. After presenting the $L=4$ canonical form, we will focus on its integration by the method of differential equations. We will show how to decompose the integrals into a basis of master integrals, which can be computed from a set of first order differential equations. We will discuss how to take those differential equations into a canonical form, which highly simplifies their analysis. Finally, we will solve the canonical differential equations and we will use their solution to compute the $L=4$ integrated negative geometries. 

\subsection{$L=4$ canonical form}

As shown in \cite{He:2022cup}, the bipartite constraint that must be imposed in \eqref{bipartite expansion} implies that only three type of negative geometries contribute to the $L=4$ canonical form $\tilde{\Omega}_4$. On the one hand, one has to consider the \textit{ladder} diagrams
$$
\includegraphics[scale=0.27]{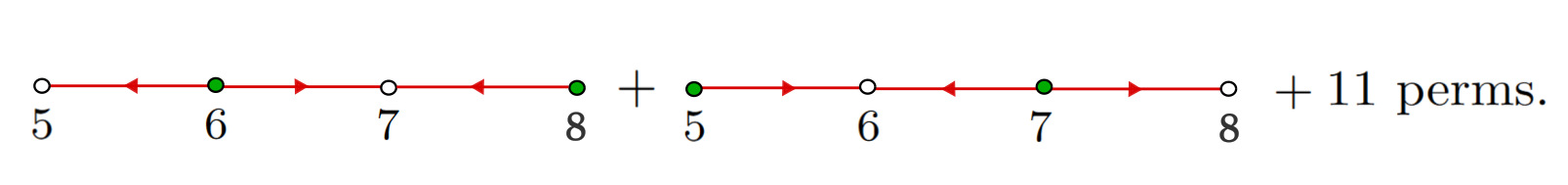}
$$
whose canonical form is
\begin{equation}
\label{ladder canonical form}
\tilde{\Omega}_{4}^{\rm Ladd} = 8 c^2 \frac{\epsilon_6\epsilon_7}{s_5 t_6 s_7 t_8 D_{56}D_{67}D_{78}} + (s \leftrightarrow t) + {\rm perms}.
\end{equation}
On the other hand, there is a contribution from the \textit{star} diagrams
\vspace{0.15cm}
$$
\includegraphics[scale=0.27]{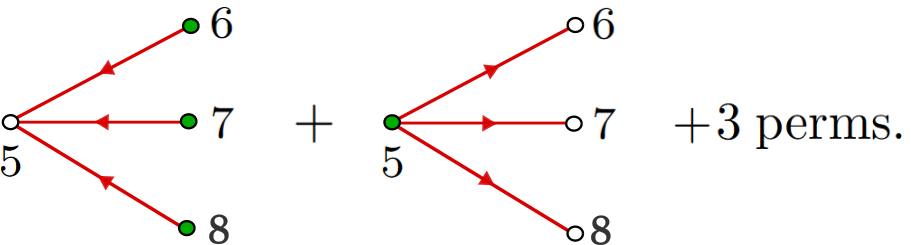}
$$
with
\begin{equation}
\label{star canonical form}
\tilde{\Omega}_{4}^{\rm Star} = 8 c^3 \frac{t_5}{s_5 t_6 t_7 t_8 D_{56}D_{57}D_{58}} + (s \leftrightarrow t) + {\rm perms}.
\end{equation}
Finally, one has to take into account also the \textit{box} diagrams
\vspace{0.15cm}
$$
\includegraphics[scale=0.27]{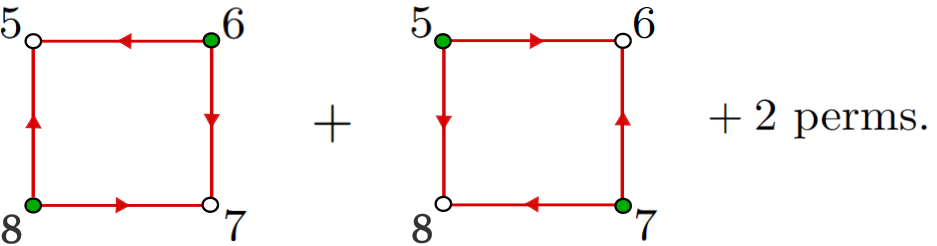}
$$
that have
\begin{equation}
\label{box canonical form}
\tilde{\Omega}_{4}^{\rm Box} =4 \frac{4 \epsilon_5 \epsilon_6 \epsilon_7 \epsilon_8-c( \epsilon_5 \epsilon_7 N^t_{68}+\epsilon_6 \epsilon_8 N^s_{57})-c^2 N^{\rm cyc}_{5,6,7,8}}{s_5 t_6 s_7 t_8 D_{56}D_{67}D_{78}D_{58}} + (s \leftrightarrow t) + {\rm perms}.
\end{equation}
and where
\begin{align}
\label{box twistor def-1}
N^s_{57} &= \langle l_5 12 \rangle \langle l_7 34 \rangle + \langle l_7 12 \rangle \langle l_5 34 \rangle \,, \\
N^t_{68} &= \langle l_6 14 \rangle \langle l_8 23 \rangle + \langle l_8 14 \rangle \langle l_6 23 \rangle \,, \\
\label{box twistor def-3}
N^{\rm cyc}_{ijkl} &= \langle l_i 12 \rangle \langle l_j 34 \rangle \langle l_k 12 \rangle \langle l_l 34 \rangle + \langle l_i 23 \rangle \langle l_j 14 \rangle \langle l_k 23 \rangle \langle l_l 14 \rangle  \,.
\end{align}
The canonical form $\tilde{\Omega}_4$ is then given as
\begin{equation}
    \label{four-loop canonical form}
    \tilde{\Omega}_4 = -\tilde{\Omega}_{4}^{\rm Ladd}-\tilde{\Omega}_{4}^{\rm Star}+\tilde{\Omega}_{4}^{\rm Box} \,.
\end{equation}

\subsection{Star diagrams}

Before turning to the computation of the integrals over $\tilde{\Omega}_4^{\rm Ladd}$ and $\tilde{\Omega}_4^{\rm Box}$ with the method of differential equations, let us discuss the integration over $\tilde{\Omega}_{4}^{\rm Star}$, which can be easily done by direct integration.

We should recall that in order to compute the ${\cal F}_3$ and ${\cal G}_3$ functions that were defined in \eqref{n=4 general expression D=3} one should freeze one of the loop integrations in ${\cal L}_4$ and then perform the remaining three-loop integrations. With the freedom of arbitrarily choosing a frozen node, we will
always leave the $X_5$ variable unintegrated. In order to diagrammatically represent this integration process we will use a squared node to denote the loop variable that is left unintegrated. For example, for the star diagrams we have two possibilities: on the one hand, one can freeze the node that is at the center of the graph, i.e.
\vspace{0.15cm}
\begin{equation}
\includegraphics[scale=0.27]{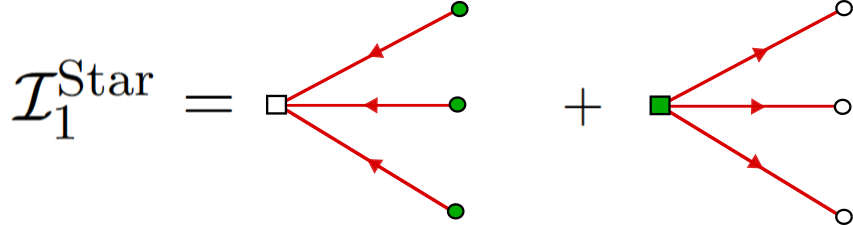}
\end{equation}
On the other hand, the unintegrated node can be in one of the legs of the star, i.e.
\vspace{0.15cm}
\begin{equation}
\includegraphics[scale=0.27]{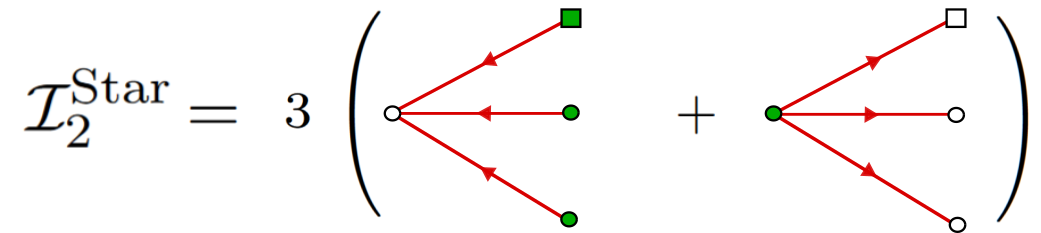}
\end{equation}
where the factor of 3 comes from the permutation of the loop variables that are being integrated. Interestingly, both the ${\cal I}_1^{\rm Star}$ and the ${\cal I}_2^{\rm Star}$ integrals can be straightforwardly solved by taking into account the triangle integral
\begin{equation}
\label{triangle integral-1}
\int \frac{d^3X_6}{i \pi^{3/2}} \, \frac{1}{X_{26}^2 X_{46}^2 X_{56}^2} = \frac{\pi^{3/2}}{\sqrt{X_{25}^2 X_{45}^2 X_{24}^2}} \,,
\end{equation}
which can be easily computed using Feynman parametrization. Then, one gets
\begin{equation}
\label{I1 Star}
\begin{aligned}
{\cal I}_1^{\rm Star} &= 8\int \frac{d^3l_6}{i \pi^{3/2}} \int \frac{d^3l_7}{i \pi^{3/2}} \int \frac{d^3l_8}{i \pi^{3/2}} \, \frac{c^3 \, t_5}{s_5 t_6 t_7 t_8 D_{56}D_{57}D_{58}} + (s \leftrightarrow t)  \\
&= -8 \pi^{9/2} \left( \frac{X^2_{13}X^2_{24}}{X^2_{15}X^2_{25}X^2_{35}X^2_{45}} \right)^{3/4} \, \left( z^{1/4}+ \frac{1}{z^{1/4}} \right) \,,
\end{aligned}
\end{equation}
and
\begin{equation}
\label{I2 Star}
\begin{aligned}
{\cal I}_2^{\rm Star} &= 24 \int \frac{d^3l_6}{i \pi^{3/2}} \int \frac{d^3l_7}{i \pi^{3/2}} \int \frac{d^3l_8}{i \pi^{3/2}} \, \frac{c^3 \, t_6}{s_6 t_5 t_7 t_8 D_{56}D_{67}D_{68}} + (s \leftrightarrow t)  \\
&= -24 \pi^{9/2} \left( \frac{X^2_{13}X^2_{24}}{X^2_{15}X^2_{25}X^2_{35}X^2_{45}} \right)^{3/4} \, \left( z^{1/4}+ \frac{1}{z^{1/4}} \right) \,.
\end{aligned}
\end{equation}
Therefore, summing up \eqref{I1 Star} and \eqref{I2 Star} and taking into account the normalization \eqref{tilde n L} (which gives $n_4=\frac{1}{4!} \left( \frac{i}{2\sqrt{\pi}}\right)^4$ at $L=4$) we arrive at
\begin{align}
\label{F and G Star}
{\cal F}_3^{\rm Star} (z) &= -\frac{\pi^2}{12} \left( z^{1/4} + \frac{1}{z^{1/4}} \right) \,, \; &{\cal G}_3^{\rm Star} (z) &=0 \,.
\end{align}

\subsection{Ladder and box diagrams}

Let us now discuss the integration of the ladder and box diagrams. For the case of the ladders we can either integrate a node at the end of the chain, i.e.
\vspace{0.15cm}
\begin{equation}
\includegraphics[scale=0.27]{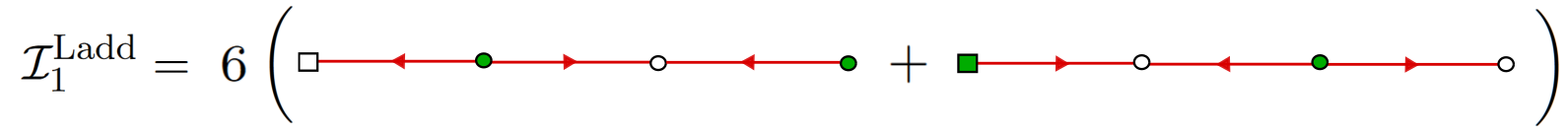}
\end{equation}
or a node in the middle of the chain, i.e.
\vspace{0.15cm}
\begin{equation}
\includegraphics[scale=0.27]{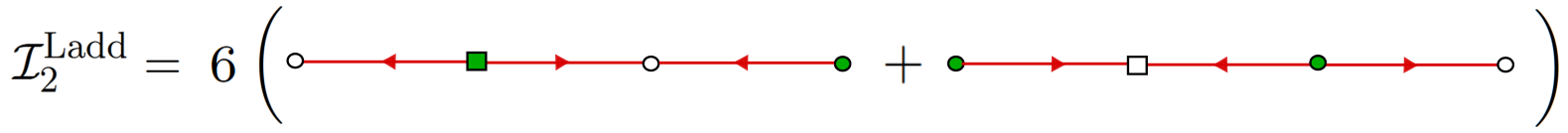}
\end{equation}
On the other hand, for the integration of the box diagram we have only one contribution, given by
\vspace{0.15cm}
\begin{equation}
\includegraphics[scale=0.27]{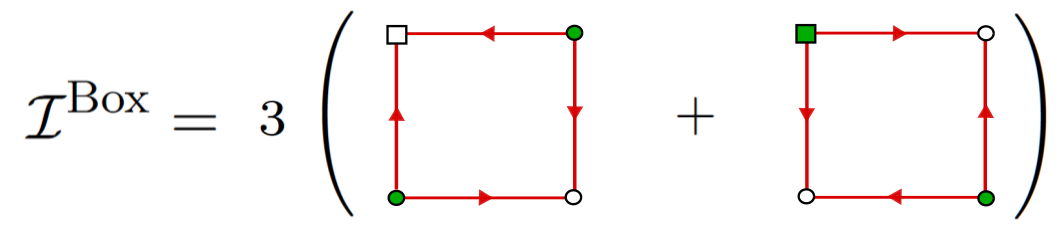}
\end{equation}
To be more specific, the integrals that we seek to compute are
\begin{align}
\label{I1 ladder}
{\cal I}_1^{\rm Ladd}&= 48 \int \frac{d^3l_6}{i \pi^{3/2}} \int \frac{d^3l_7}{i \pi^{3/2}} \int \frac{d^3l_8}{i \pi^{3/2}} \, \frac{c^2 \epsilon_6\epsilon_7}{s_5 t_6 s_7 t_8 D_{56}D_{67}D_{78}} + (s \leftrightarrow t) \,, \\
\label{I2 ladder}
{\cal I}_2^{\rm Ladd}&= 48 \int \frac{d^3l_6}{i \pi^{3/2}} \int \frac{d^3l_7}{i \pi^{3/2}} \int \frac{d^3l_8}{i \pi^{3/2}} \, \frac{c^2 \epsilon_5\epsilon_7}{t_6 s_5 t_7 s_8 D_{56}D_{57}D_{78}} + (s \leftrightarrow t) \,, \\
\label{I box}
{\cal I}^{\rm Box}&= 12 \int \frac{d^3l_6}{i \pi^{3/2}} \int \frac{d^3l_7}{i \pi^{3/2}} \int \frac{d^3l_8}{i \pi^{3/2}} \, \frac{4 \epsilon_5 \epsilon_6 \epsilon_7 \epsilon_8-c( \epsilon_5 \epsilon_7 N^t_{68}+\epsilon_6 \epsilon_8 N^s_{57})-c^2 N^{\rm cyc}_{5,6,7,8}}{s_5 t_6 s_7 t_8 D_{56}D_{67}D_{78}D_{58}} + (s \leftrightarrow t) \,.
\end{align}

\subsubsection{Differential equations}
\label{sec master integrals}

In the remaining of this section we will discuss the computation of \eqref{I1 ladder}-\eqref{I box}
by the method of differential equations \cite{Henn:2013pwa}.
As the first step towards that goal, let us begin by defining the family of integrals given as
\begin{equation}
\label{three loop family}
G_{a_1, \, a_2, \, a_3, \, \dots , \, a_{15}}= \int \frac{d^D l_1}{i\pi^{3/2}} \, \int \frac{d^D l_2}{i\pi^{3/2}} \, \int \frac{d^D l_3}{i\pi^{3/2}} \, \prod_{j=1}^{15} \frac{1}{P_j^{a_j}} \,,
\end{equation}
with $D=3-2\epsilon$ and
\begin{center}
\begin{tabular}{l l l}
$P_1 = -(l_1+p_1)^2 \,,$ & $P_6 = -(l_3-p_4)^2 \,,$ & $P_{11} = -(l_2+p_1)^2 \,,$ \\
$P_2 = -(l_1-p_4)^2 \,,$ & $P_7 = -(l_1-l_2)^2 \,,$ & $P_{12} = -(l_2-p_4)^2 \,,$\\
$P_3 = -l_2^2 \,,$ & $P_8 = -(l_2-l_3)^2 \,,$ & $P_{13} = -l_3^2 \,,$ \\
$P_4 = -(l_2+p_1+p_2)^2 \,,$ & $P_9 = -l_1^2 \,,$ & $P_{14} = -(l_3+p_1+p_2)^2 \,,$\\
$P_5 = -(l_3+p_1)^2 \,,$ & $P_{10} = -(l_1+p_1+p_2)^2 \,,$ & $P_{15} = -(l_1-l_3)^2 \,,$
\end{tabular}
\end{center}
and where the momenta $p_i,\, i=1 , \, \dots , \, 4$ are all massless. In order to express \eqref{I1 ladder}-\eqref{I box} as a linear combination of integrals of the family \eqref{three loop family} we use the identity
\begin{equation}
\label{5D epsilon product}
\epsilon(1,2,3,4,i) \, \epsilon(1,2,3,4,j)= \frac{X^4_{13} X^4_{24}}{32} \left( \frac{X^2_{1i} X^2_{3j}+X^2_{1j} X^2_{3i}}{X^2_{13}} + \frac{X^2_{2i} X^2_{4j}+X^2_{2j} X^2_{4i}}{X^2_{24}} - X^2_{ij}\right) \,,
\end{equation}
which allows us to rewrite \eqref{I1 ladder}-\eqref{I box} in terms of integrals that only depend on $X_{ij}^2$ distances. However, let us note that these integrals still do not belong to the family defined in \eqref{three loop family}, given that there is still a non-trivial dependence on $x_5$. One can get rid of that dependence by taking into account the dual conformal covariance of the integrals \eqref{I1 ladder}-\eqref{I box}, that allows one to go to the $x_5 \to \infty$ frame. Therefore, we define
\begin{align}
\label{I1 ladder infty}
\hat{\cal I}_1^{\rm Ladd} &= \lim_{x_5 \to \infty} (x_5^2)^3 \, {\cal I}_1^{\rm Ladd} \,, \\
\label{I2 ladder infty}
\hat{\cal I}_2^{\rm Ladd} &= \lim_{x_5 \to \infty} (x_5^2)^3 \, {\cal I}_2^{\rm Ladd} \,, \\
\label{I box infty}
\hat{\cal I}^{\rm Box} &= \lim_{x_5 \to \infty} (x_5^2)^3 \, {\cal I}^{\rm Box} \,.
\end{align}
The advantage of taking the $x_5 \to \infty$ limit relies in that it allows us to express the integrals \eqref{I1 ladder infty}-\eqref{I box infty} solely in terms of four-particle kinematic variables, i.e. in that limit one can rewrite \eqref{I1 ladder infty}-\eqref{I box infty} as a linear combination of integrals of the type defined in \eqref{three loop family}.
Moreover, we can further decompose \eqref{I1 ladder infty}-\eqref{I box infty} in terms of a set of master integrals for the family \eqref{three loop family}. In order to look for that basis we have used the \texttt{FIRE} \cite{Smirnov:2023yhb} and \texttt{LiteRed} \cite{Lee:2013mka} algorithms, that have allowed us to find the following basis of 19 master integrals:
\begin{align}
    M_1 &= s^{-1/2 + 3 \epsilon} \, \hat{G}_{0, 1, 0, 0, 1, 0, 1, 1} \,, &M_2 &= s^{3/2 + 3 \epsilon} \, \hat{G}_{0, 1, 0, 1, 1, 1, 1, 1}\,, \nonumber \\
    M_3 &= s^{3/2 + 3 \epsilon}\, \hat{G}_{0, 1, 1, 1, 0, 1, 1, 1}\,,  &M_4 &= s^{3/2 + 3 \epsilon} \, \hat{G}_{0, 1, 1, 1, 1, 0, 1, 1} \,, \nonumber\\
    M_5 &= s^{5/2 + 3 \epsilon}\, \hat{G}_{0, 1, 1, 1, 1, 0, 1, 2}\,,  &M_6 &= s^{3/2 + 3 \epsilon} \, \hat{G}_{0, 1, 1, 1, 1, 1, 1, 0}\,, \nonumber\\
    M_7 &= s^{5/2 + 3 \epsilon}\, \hat{G}_{0, 1, 1, 1, 1, 1, 1, 1} \,, &M_8 &= s^{7/2 + 3 \epsilon} \, \hat{G}_{0, 1, 1, 1, 1, 1, 1, 2}\,, \nonumber\\
    \label{non can MI}
    M_9 &= s^{7/2 + 3 \epsilon}\, \hat{G}_{0, 1, 1, 1, 1, 1, 2, 1}\,,  &M_{10} &= s^{3/2 + 3 \epsilon} \, \hat{G}_{1, 1, 0, 0, 1, 1, 1, 1} \,, \\
    M_{11} &= s^{3/2 + 3 \epsilon}\, \hat{G}_{1, 1, 0, 1, 1, 1, 0, 1} \,, &M_{12} &= s^{5/2 + 3 \epsilon} \, \hat{G}_{1, 1, 0, 1, 1, 1, 1, 1}\,,\nonumber \\
    M_{13} &= s^{3/2 + 3 \epsilon}\, \hat{G}_{1, 1, 1, 1, 1, 1, 0, 0} \,, &M_{14} &= s^{5/2 + 3 \epsilon} \, \hat{G}_{1, 1, 1, 1, 1, 1, 0, 1}\,, \nonumber\\
    M_{15} &= s^{7/2 + 3 \epsilon}\, \hat{G}_{1, 1, 1, 1, 1, 1, 0, 2}\,,  &M_{16} &= s^{7/2 + 3 \epsilon} \, \hat{G}_{1, 1, 1, 1, 1, 1, 1, 1}\,,\nonumber \\
    M_{17} &= s^{9/2 + 3 \epsilon}\, \hat{G}_{1, 1, 1, 1, 1, 1, 1, 2}  \,,&M_{18} &= s^{7/2 + 3 \epsilon} \, \hat{G}_{1, 1, 1, 1, 1, 2, 0, 1} \,,\nonumber \\
    M_{19} &= s^{9/2 + 3 \epsilon}\, \hat{G}_{1, 1, 1, 1, 1, 2, 1, 1} \,, &  & \nonumber
\end{align}
where we are using the shorthand notation
\begin{equation}
    \hat{G}_{a_1, a_2, a_3, a_4, a_5, a_6, a_7, a_8} :=G_{a_1, a_2, a_3, a_4, a_5, a_6, a_7, a_8, 0,0,0,0,0,0,0} \,.
\end{equation}
Let us note that we have choosen to normalize each element of the basis with a factor
\begin{equation}
    \label{3L normalization}
    s^{-\frac{9}{2}+3 \epsilon+ \sum_{i=1}^{15} a_i} \,,
\end{equation}
in order to get dimensionless integrals.

As it is well known, for a given basis of master integrals one can generically derive a set of first order differential equations that formally allows for their computation. For the case of interest to us, we have
\begin{equation}
\label{non canonical diff eq}
\partial_{\omega} \vec{M} = {\bf A}(\omega,\epsilon) \, \vec{M} \,,
\end{equation}
where $\vec{M}$ is a vector constructed with the 19 master integrals presented in \eqref{non can MI}, ${\bf A}$ is a $19\times19$ matrix presented in the file {\tt non\_can\_DE.txt}, $\epsilon$ is the dimensional regularization parameter (defined as $D=3-2\epsilon$), and $\omega$ is defined by\footnote{For each value of $t/s$ there are four possible solutions of \eqref{omega}. Throughout this paper we will always choose to work with the solution that lies in the $0< \omega <1$ interval.}
\begin{equation}
\label{omega}
\frac{t}{s}=\frac{1}{4} \left( \omega - \frac{1}{\omega} \right)^2 \,,
\end{equation}
where the Mandelstam variables are taken to be $s=(p_1+p_2)^2$ and $t=(p_1+p_4)^2$. The reason for introducing the $\omega$ variable defined in \eqref{omega} will become clear soon. 

As shown in \cite{Henn:2013pwa}, in order to solve \eqref{non canonical diff eq} it is better to look for a change of basis 
\begin{equation}
\label{change of basis}
\vec{M}= {\bf T} \, \vec{C} \,,
\end{equation}
that takes the differential equation \eqref{non canonical diff eq} into a \textit{canonical form}
\begin{equation}
\label{canonical diff eq}
\partial_{\omega} \vec{C} = \epsilon \, {\bf A}_c(\omega) \, \vec{C} \,,
\end{equation}
where the matrix ${\bf A}_c(\omega)$ is given as
\begin{equation}
\label{canonical matrix}
{\bf A}_c(\omega) = \sum_{j=1}^{n} \, {\bf a}_j \, {\rm dlog} [W_j(\omega)] \,,
\end{equation}
with constant ${\bf a}_j$ matrices and where the $W_j$ are a set of algebraic functions of $\omega$. The $W_j$ functions are known as \textit{letters}, and the set of all letters is known as the \textit{alphabet}. Then, expanding the canonical basis as\footnote{In order to get \eqref{C epsilon expansion} one has to properly normalize the canonical basis so that all integrals are finite in $\epsilon \to 0$. }
\begin{equation}
\label{C epsilon expansion}
\vec{C}(\omega, \epsilon)=\sum_{k=0}^{\infty} \epsilon^k \, \vec{C}^{(k)} (\omega) \,,
\end{equation}
one gets
\begin{equation}
\vec{C}^{(0)} (\omega)= {\rm const.} \,,
\end{equation}
and
\begin{equation}
\label{formal solution canonical DE}
\vec{C}^{(k)} (\omega)=\vec{C}^{(k)} (0)+ \int_{0}^{\infty} d\tau \, {A}_c(\tau) \, \vec{C}^{(k-1)} (\tau) \,,
\end{equation}
for $k \geq 1$, where $\vec{C}^{(k)} (0)$ are boundary constants at the boundary point $\omega=0$\footnote{The choice of the boundary point is arbitrary.}. Therefore, from \eqref{canonical matrix} and \eqref{formal solution canonical DE} one can see that the solution to the differential equation \eqref{canonical diff eq} will be a set of integrals with uniform transcendental weight in their $\epsilon$ expansion, provided the $\vec{C}^{(0)}$ vector has a given transcendental weight $w_0$.

In order to take the differential equation \eqref{non canonical diff eq} into the canonical form we have used the algorithm \texttt{CANONICA} \cite{Meyer:2017joq}. We have found that, given the change of basis presented in the file {\tt change\_of\_basis.txt} (which corresponds to the $\bf{T}$ matrix in \eqref{change of basis}), the alphabet is
\begin{align}
W_1 &= \omega \,, \qquad & W_2 &=1+\omega \,, \\
W_3 &= 1-\omega \,, \qquad & W_4 &=1+\omega^2 \,.
\end{align}
The corresponding ${\bf a}_j$ matrices are presented in the ancillary files ${\tt a1}.{\tt txt}$, ${\tt a2}.{\tt txt}$, ${\tt a3}.{\tt txt}$ and ${\tt a4}.{\tt txt}$. At this point we are in the position of arguing the choice of $\omega$ as the kinematic variable to construct the differential equations. Had we used $z=t/s$ instead of $\omega$, we would not have got a rational alphabet, and the letters would have included square roots of $z$. The decompositions of \eqref{I1 ladder infty}-\eqref{I box infty} into the canonical basis $\vec{C}$ are presented in the ancillary files ${\tt I1} \_ {\tt ladd} \_ {\tt dec}.{\tt txt}$, ${\tt I2} \_ {\tt ladd} \_ {\tt dec}.{\tt txt}$ and ${\tt I} \_ {\tt box} \_ {\tt dec}.{\tt txt}$.

\subsubsection{Solving the differential equations}

As discussed in the previous section, given a set of canonical differential equations, their solution reduces to the process of finding the boundary values $\vec{C}^{(k)}(0), \, k \geq 0$. As we expect the ${\cal F}_3$ and ${\cal G}_3$ functions to have transcendental weight $2$, we will only be interested in solving the differential equations up to second order in the $\epsilon$ expansion. Therefore, we only need to compute $\vec{C}^{(0)}(0), \, \vec{C}^{(1)}(0)$ and $\vec{C}^{(2)}(0)$. To that aim we take into consideration that
\begin{enumerate}
    \item Given that we are working with planar integrals, we should not have discontinuities for real negative values of $z=t/s$. Therefore, the symbol \cite{Goncharov:2010jf} of the  $\vec{C}$ integrals should always have $z=\frac{(1-\omega^2)^2}{4 \omega^2}$ as its first letter. Moreover, there should not be singularities for $u \to 0$, which taking into account the pole structure of the ${\bf A}_c$ matrix implies
\begin{equation}
    \label{two loop sing constraint}
    \lim_{\omega^2 \to -1} {\bf a}_4 \, \vec{C}^{(k)} (\omega)=0 \,.
\end{equation}
for all $k \geq 0$.

    \item Four out of the nineteen integrals in the canonical basis can be computed exactly by Feynman parametrization \cite{Brandhuber:2013gda}. More precisely, we have 
    \begin{align}
    \label{three loop input data 1}
        C_1 (z) &= -z^{-3\epsilon} \, \frac{9 \left(24 \epsilon ^2-10 \epsilon +1\right) \Gamma \left(\frac{1}{2}-\epsilon \right)^4 \Gamma \left(3 \epsilon -\frac{1}{2}\right)}{10 \epsilon ^3 (2 \epsilon +1)^3 \Gamma (2-4 \epsilon )} \,, \\
        C_6 (z) &= z^{-\epsilon} \, \frac{54 \Gamma \left(\frac{1}{2}-\epsilon \right)^4 \Gamma (-2 \epsilon ) \Gamma \left(\epsilon +\frac{1}{2}\right) \Gamma \left(\epsilon +\frac{3}{2}\right) \Gamma (2 \epsilon +2)}{\epsilon ^2 (2 \epsilon +1)^5 \Gamma \left(\frac{1}{2}-3 \epsilon \right) \Gamma (1-2 \epsilon )} \,, \\
            \label{three loop input data 3}
        C_{11} (z) &=z^{-3\epsilon} \, \frac{216 \Gamma \left(\frac{1}{2}-\epsilon \right)^4 \Gamma (-2 \epsilon ) \Gamma \left(\epsilon +\frac{1}{2}\right) \Gamma \left(\epsilon +\frac{3}{2}\right) \Gamma (2 \epsilon +2)}{\epsilon ^2 (2 \epsilon +1)^5 \Gamma \left(\frac{1}{2}-3 \epsilon \right) \Gamma (1-2 \epsilon )} \,, \\
             \label{three loop input data 4}
        C_{13} (z) &=z^{-2\epsilon} \, \frac{108 \Gamma \left(\frac{1}{2}-\epsilon \right)^6 \Gamma \left(\epsilon +\frac{1}{2}\right)^3}{\epsilon  (2 \epsilon +1)^3 \Gamma (1-2 \epsilon )^3} \,.
    \end{align}
Consequently, the solutions obtained from \eqref{canonical diff eq} should agree with \eqref{three loop input data 1}-\eqref{three loop input data 4}. 

\item Within the euclidean region, which in our conventions\footnote{We are taking the metric to have signature $(-,+,+)$ and we are defining $s:=(p_1+p_2)^2$ and $t:=(p_1+p_4)^2$.} is defined by the $s>0$ and $t>0$ constraints (or, equivalently, $0<\omega<1$), all the integrals in the $\vec{C}$ basis should be real. 

\end{enumerate}

We have used the \texttt{PolyLogTools} package \cite{Duhr:2019tlz} to manipulate the iterated integrals that appear when solving the differential equations. All the above considerations allow us to completely fix the zeroth-order and first-order constants $\vec{C}^{(0)}$ and $\vec{C}^{(1)}$. As for the second-order boundary vector $\vec{C}^{(2)}$, the previous constraints leave only two unknown constants. We have fixed them by demanding consistence of the $\vec{C}$ integrals with their numerical evaluations. The $\vec{C}$ canonical basis is presented, up to second order in its $\epsilon$-expansion, in the ancillary file ${\tt DE} \_ {\tt sol}.{\tt txt}$.

\subsubsection{Integrated results}
\label{subsec Three-loop integrated negative geometry}

The solution to the canonical differential equations discussed in previous sections allows us to compute the ladder and box integrals defined in \eqref{I1 ladder infty}-\eqref{I box infty}. We get
\begin{align}
\label{I1 ladder infty-result}
\hat{\cal I}_1^{\rm Ladd} &= -24 \, s \sqrt{t} \, \pi^{5/2}   \left[ \left( \log(z)-2 \log(2) \right)^2 + 2\pi^2 \right]   \nonumber \\
    & \quad - 24 \, t \sqrt{s} \, \pi^{5/2}   \left[  \left( \log(z)+2 \log(2) \right)^2 +2 \pi^2 \right] \,, \\
\label{I2 ladder infty-result}
\hat{\cal I}_2^{\rm Ladd} &= -48 \pi^{5/2} \sqrt{s t (s+t)} \left[ {\cal H}(z) + {\cal H} \left( \frac{1}{z} \right) + 2 \pi^2 \right] \,, \\
\label{I box infty-result}
\hat{\cal I}^{\rm Box} &= -48 \, \pi^{5/2} \,\sqrt{s t (s+t)} \, \left[ {\cal H}(z) + {\cal H} \left( \frac{1}{z} \right) \right]  \nonumber \\
&\quad -32\pi^{9/2} \,\left( 3\, \sqrt{s t (s+t)}- \, s \sqrt{t}-\, t \sqrt{s}
  \right) \,,
\end{align}
where the ${\cal H}$ function was defined in \eqref{H-app}. Let us note that the ${\cal I}_2^{\rm Ladd}$ integral can also be easily obtained from direct integration using the two-loop integrals computed in \cite{Henn:2023pkc,He:2023exb}, in agreement with the result presented in \eqref{I2 ladder infty-result}. Reversing the $x_5 \to \infty$ limit introduced in \eqref{I1 ladder infty}-\eqref{I box infty} and taking into account the normalizations presented in \eqref{normalization L_L} we arrive at
\begin{align}
\label{F Ladder}
{\cal F}_3^{\rm Ladd} (z) &= -\frac{z^{-1/4}}{16} \,  \left[ \left( \log(z)-2 \log(2) \right)^2 + 2\pi^2 \right]   \nonumber \\
&\quad - \frac{z^{1/4}}{16} \,   \left[ \left( \log(z)+2 \log(2) \right)^2 + 2\pi^2 \right]  \nonumber \\
&\quad -\frac{z^{-1/4} \sqrt{1+z}}{8} \left[  {\cal H}(z)+  {\cal H} \left( \frac{1}{z} \right) + 2 \pi^2 \right] \\
\label{F Box}
{\cal F}_3^{\rm Box} (z) &= -\frac{z^{-1/4}\sqrt{1+z}}{8} \, \left[ {\cal H} (z) + {\cal H} \left( \frac{1}{z} \right) \right]  \nonumber \\
&\quad -\frac{\pi^2}{12} \left( 3 z^{-1/4}\sqrt{1+z}-z^{1/4} - z^{-1/4} \right) \,,
\end{align}
and
\begin{align}
\label{G Ladd and Box}
{\cal G}_3^{\rm Ladd} (z)= {\cal G}_3^{\rm Box} (z)  &= 0 \,.
\end{align}
Finally, adding up the contributions from the different diagrams as in \eqref{four-loop canonical form}, we get
\begin{align}
{\cal F}_3 (z) &= -{\cal F}_3^{\rm Ladd} (z)-{\cal F}_3^{\rm Star} (z)+{\cal F}_3^{\rm Box} (z) \\
\label{F total}
&= \frac{z^{-1/4} }{16} \, \left[ \log(z)-2 \log(2) \right]^2 + \frac{z^{1/4}}{16} \,   \left[ \log(z)+2 \log(2) \right]^2 \nonumber \\
&\quad + \frac{7\pi^2}{24} \left( z^{1/4} + z^{-1/4}  \right)  \,, \\
\label{G total}
{\cal G}_3 (z) &=0 \,.
\end{align}
We note that the above $L=4$ formulas, together with the $L \leq 3$ results presented in \eqref{tree level F and G}-\eqref{two loop F and G}, suggest that
\begin{align}
 {\cal F}_{2k+1} (z) \text{ and } {\cal G}_{2k} (z)  \text{ have transcendental degree $2k$}
\end{align}
for $k \geq 0$.

\section{Properties of the $L=4$ result}
\label{sec properties}

After computing the $L=4$ integrated negative geometries of the ABJM theory, we will turn now to the analysis of the integrated results. We will begin with a discussion on their leading singularities, and then we will move to an analysis of the sign patterns that are observed in the integrated results.

\subsection{Leading singularities}

One can always generically write the integrated negative geometries as
\begin{equation}
\label{eq:leading_singularities_def}
\left( \prod_{j=6}^{4+L} \int \frac{d^3 x_j}{i\pi^{3/2}} \right) \, {\cal L}_L = \sum_{i=1}^k R_{L-1,i}\, T_{L-1,i}\,,
\end{equation}
for some integer $k$, where the $T_{L-1,i}$ are transcendental functions and the $R_{L-1,i}$ are rational functions known as \textit{leading singularities}. 
For the sake of more generality, we will separately study the contributions of each individual negative geometry diagram, without considering the cancellations that may occur when adding up all contributions according to \eqref{bipartite expansion}. Taking into account the $L \leq 3$ results of \cite{Henn:2023pkc,He:2023exb} and the $L=4$ results presented in the previous section, we see that the only leading singularities that contribute to the integrated results are\footnote{The $R_6$ leading singularity appears when integrating the box and ladder diagrams at $L=4$, but it cancels when summing all contributions as in \eqref{bipartite expansion}.}
\begin{align}
\label{even leading sing}
R_{1} &= \frac{4 \, \epsilon(1,2,3,4,5)}{X_{15}^2X_{25}^2X_{35}^2X_{45}^2} \,, & R_{2} &= \frac{R_{1}}{\sqrt{1+z}} \,, \quad & R_{3} &= \sqrt{\frac{z}{z+1}} \, R_{1} \,, \\
R_{4} &= \left( \frac{X_{13}^2 X_{24}^2}{X_{15}^2 X_{25}^2 X_{35}^2 X_{45}^2} \right)^{3/4} \,  z^{-1/4}\, \,, & R_{5} &= \sqrt{z} \, R_4 \,, & R_6&= \sqrt{1+z} \, R_4 \,.
\end{align}
As discussed in \cite{Henn:2023pkc}, in order to work only with four-particle kinematic variables it is best to perform the analysis of leading singularities in the frame in which the unintegrated variable goes to infinity. Therefore, defining
\begin{align}
\label{x5 infty LS}
r= \lim_{x_5 \to \infty} (x_5^2)^3 \, R \,,
\end{align}
we see that in the $x_5 \to \infty$ limit the leading singularities reduce to
\begin{align}
r_1 &= \sqrt{s \, t \, (s+t)} \,, \qquad &r_2 &= s \sqrt{t} \,,  \qquad &r_3 &= t \sqrt{s} \,, \\
r_4 &= s \sqrt{t} \,, \qquad &r_5 &= t \sqrt{s} \,, \qquad &r_6 &= \sqrt{s \, t \, (s+t)} \,. \\
\end{align}
That is, we get that up to $L \leq 4$ the leading singularities of each individual integrated negative geometry diagram belong to the set 
\begin{equation}
\{ \sqrt{st(s+t)} , \, s\sqrt{t}, \, t\sqrt{s}\} \,.
\end{equation}

Finally, let us comment about the conformal symmetry of the integrated results. In \cite{Henn:2023pkc}, it was shown that the leading singularities of the $L \leq 3$ integrated results are conformally invariant in the $x_5 \to \infty$ limit and after normalization with a three-dimensional generalization of the Parke-Taylor factor. We should note that this result is in fact general for every arbitrary function of $s$ and $t$ with dimension $1$ in energy units. As a corollary, the complete integrated results (i.e. including also the transcendental functions) are conformally invariant in the $x_5 \to \infty$ limit and after normalization with the three-dimensional Parke-Taylor factor (or any other factor of dimension $-2$ in energy units). We refer to Appendix \ref{app conformal invariance} for a detailed analysis of this statement.

\subsection{Sign patterns}

We will close this section with a positivity analysis of the integrated negative geometries. To that end, let us first comment on the sign properties that have been observed for the integrated results of the ${\cal N}=4$ sYM theory. As discussed in \cite{Alday:2011ga}, dual conformal symmetry constrains the four-particle integrated negative geometries of ${\cal N}=4$ sYM  to have the form
\begin{equation}
\label{n=4 general expression}
\left( \prod_{j=6}^{4+L} \int \frac{d^4 x_j}{i\pi^{2}} \right) \, {\cal L}_L = \frac{x_{13}^2 x_{24}^2}{x_{15}^2 x_{25}^2 x_{35}^2 x_{45}^2} \, \frac{{\cal F}_{L-1} \left( z \right)}{\pi^2} \,,
\end{equation}
where $z$ is defined as in \eqref{cross ratio}. From an inspection of the $L \leq 3$ results, it was suggested in \cite{Arkani-Hamed:2021iya}
\begin{align}
{\cal F}_{L-1}(z)&<0  \qquad \text{for odd }L \,, \\
{\cal F}_{L-1}(z)&>0  \qquad \text{for even }L \,,
\end{align}
when restricted to the Euclidean region $z>0$. Moreover, similar results were found for the five-particle case in \cite{Chicherin:2022zxo}. There it was found that
\begin{equation}
\label{five particle signs}
    f_5^{(0)} \big|_{\rm Eucl^+}>0 \,, \qquad f_5^{(1)} \big|_{\rm Eucl^+}<0 \,, \qquad f_5^{(2)} \big|_{\rm Eucl^+}>0 \,,
\end{equation}
where
\begin{equation}
f_5^{(L-1)}=\lim_{x_6 \to \infty}  \, \left( \prod_{j=7}^{5+L} \int \frac{d^4 x_j}{i\pi^{2}} \right) \, {\cal L}_L^{(n=5)} \,,
\end{equation}
with ${\cal L}_L^{(n=5)}$ the five-particle $L$-loop integrand for the logarithm of the amplitude. Let us note that $x_5$ is no longer an unintegrated loop variable in \eqref{five particle signs}, and such a role is played by $x_6$ in that case. Furthermore, the region ${\rm Eucl^+}$ in \eqref{five particle signs} is defined as the one were all adjacent five-particle Mandelstam invariants are negative and
\begin{equation}
4i \epsilon_{\mu \nu \rho \sigma} p_1^{\mu} p_2^{\nu} p_3^{\rho} p_4^{\sigma} >0 \,.
\end{equation}
Let us turn now to the analysis of the ABJM case. For the $L \leq 3$ the results reviewed in \eqref{tree level F and G}-\eqref{two loop F and G} one can read that in the Euclidean region, i.e. for $z>0$,
\begin{align}
{\cal G}_0(z)\big|_{z>0}<0 \,, \qquad {\cal F}_1(z)\big|_{z>0}<0 \,, \qquad {\cal G}_2(z)\big|_{z>0}>0 \,.
\end{align}
Moreover, from Section \ref{sec three loop results} we see that in the $L=4$ case the individual negative geometry diagrams behave as
\begin{align}
{\cal F}_3^{\rm Ladd}(z)\big|_{z>0}<0 \,, \qquad {\cal F}_3^{\rm Star}(z)\big|_{z>0}<0 \,, \qquad {\cal F}_3^{\rm Box}(z)\big|_{z>0}>0 \,,
\end{align}
and, therefore, from \eqref{four-loop canonical form} we get
\begin{align}
{\cal F}_3(z)\big|_{z>0}>0  \,.
\end{align}
Consequently, we conjecture that
\begin{align}
{\cal F}_{2k+1}(z)\big|_{z>0}&<0 \,, \qquad \text{and} \qquad {\cal G}_{2k}(z)\big|_{z>0}<0 \,, \qquad \text{for even }k\,,\\
{\cal F}_{2k+1}(z)\big|_{z>0}&>0 \,, \qquad \text{and} \qquad {\cal G}_{2k}(z)\big|_{z>0}>0 \,, \qquad \text{for odd } k\,,
\end{align}
for $k \geq 0$.

\section{Cusp anomalous dimension}
\label{sec gamma cusp}

Wilson loops with light-like cusps have divergences that can not be renormalized by a redefinition of the couplings of the theory \cite{Polyakov:1980ca,Dotsenko:1979wb,Brandt:1981kf,Korchemskaya:1992je}. Instead, to regularize the expectation value of cusped Wilson loops one has to introduce a $Z_{\rm cusp}$ renormalization factor, which as usual allows to define a cusp anomalous dimension as
\begin{equation}
\label{Gamma cusp definition}
\Gamma_{\rm cusp}= \mu \frac{d \log  Z_{\rm cusp}}{d\mu} \,,
\end{equation} 
where $\mu$ is the renormalization scale of the theory. The appearance of the cusp anomalous dimension is ubiquitous in Quantum Field Theory, as it does not only control the UV divergences of light-like Wilson loops, but also governs the IR divergences of scattering amplitudes \cite{Korchemsky:1991zp}. In the context of the AdS/CFT correspondence this can be seen as a consequence of the Wilson loops/scattering amplitudes duality \cite{Alday:2007hr,Drummond:2007aua,Brandhuber:2007yx,Drummond:2007cf,Henn:2010ps,Bianchi:2011dg,Chen:2011vv,Leoni:2015zxa}, that relates amplitudes with polygonal light-like Wilson loops.

Based on a proposal for the all-loop asymptotic Bethe Ansatz that describes the anomalous dimensions of large-charge single-trace operators in the ABJM theory \cite{Gromov:2008qe,Ahn:2008aa}, it was stated in \cite{Gromov:2008qe} that the cusp anomalous dimension of ABJM is related to its ${\cal N}=4$ sYM counterpart by
\begin{equation}
\label{Gamma cusp proposal ABJM}
 \Gamma_{\rm cusp}^{\rm ABJM}= \frac{1}{4} \Gamma_{\rm cusp}^{{\cal N}=4}\bigg|_{h^{{\cal N}=4} \to h^{\rm ABJM}} \,.
\end{equation}
where $h^{{\cal N}=4}$ and $h^{\rm ABJM}$ are the interpolating functions that appear in the dispersion relation of magnons in ${\cal N}=4$ sYM and ABJM, respectively \cite{Beisert:2006qh,Minahan:2008hf,Minahan:2009te,Bak:2008cp,Gaiotto:2008cg}. The $h^{{\cal N}=4}$ function was shown to be
\begin{equation}
\label{h n=4}
h^{{\cal N}=4}(\lambda)=\frac{\sqrt{\lambda}}{4\pi} \,,
\end{equation}
to all loops \cite{Gromov:2012eu}. On the other hand, for the ABJM case an all-loop conjecture was made in \cite{Gromov:2014eha} \footnote{ See \cite{Cavaglia:2016ide} for a generalization to the ABJ theory. }, where it was proposed that
\begin{equation}
\label{h conjecture ABJM}
\lambda= \frac{\sinh(2 \pi h^{\rm ABJM})}{2\pi} \, _3F_2 \left( \frac{1}{2},\frac{1}{2},\frac{1}{2} ;1, \frac{3}{2}; - \sinh^2 (2\pi h^{\rm ABJM}) \right) \,.
\end{equation}
which is in perfect agreement with the four-loop results for $h^2$ of \cite{Minahan:2009aq,Minahan:2009wg,Leoni:2010tb}. Combining \eqref{Gamma cusp proposal ABJM}, \eqref{h n=4} and \eqref{h conjecture ABJM}, one gets
\begin{equation}
\label{Gamma cusp conjecture ABJM-weak coupling-2}
\Gamma_{\rm cusp} (\lambda)= \lambda^2 - \pi^2 \lambda^4 + \frac{49 \pi^4}{30} \lambda^6 + \dots \,,
\end{equation}
for the ABJM theory, which agrees with the leading order computations made in \cite{Griguolo:2012iq,Henn:2023pkc,He:2023exb}.

As discussed in \cite{Arkani-Hamed:2021iya,Alday:2013ip,Henn:2019swt} for the ${\cal N}=4$ sYM theory and in \cite{Henn:2023pkc,He:2023exb} for the ABJM case, one of the advantages of computing integrated negative geometries is that the cusp anomalous dimension of the corresponding theory can be read from them by applying a certain functional on the integrated results. More precisely, for the ABJM theory it was shown in \cite{Henn:2023pkc,He:2023exb} that the $L$-loop contribution $\Gamma_{\rm cusp}^{(L)}$ to the cusp anomalous dimensions can be computed from the integrated negative geometries ${\cal F}_{L-1}(z)$ as
\begin{equation}
\label{Gamma cusp from F and G-functionals}
\Gamma_{\rm cusp}^{(L)} = I_{\cal F}[{\cal F}_{L-1}] \,,
\end{equation}
where
\begin{align}
\label{F functional}
I_{\cal F}[{\cal F}_{L-1}] &:= \left[-\frac{\sqrt{\pi}}{2} \int  \frac{d^D X_5}{i\pi^{D/2}} \, \left( \frac{X_{13}^2 X_{24}^2}{X_{15}^2 X_{25}^2 X_{35}^2 X_{45}^2} \right)^{\frac{3}{4}} {\cal F}_{L-1} \left( z \right) \right]_{1/\epsilon^2 \, {\rm term}} \,,
\end{align}
In particular, using 
\begin{align}
\label{F functional-zp}
I_{\cal F}[z^p] &= -\frac{2 \sqrt{\pi}}{\Gamma \left( \frac{3}{4} +p \right) \Gamma \left( \frac{3}{4} -p \right)} \,,
\end{align}
it was proved in \cite{Henn:2023pkc,He:2023exb} that the $L \leq 3$ integrated results (see \eqref{tree level F and G}-\eqref{two loop F and G}) give
\begin{align}
I_{\cal F}[{\cal F}_0] &=0 \,, \qquad &I_{\cal F}[{\cal F}_1] &=1 \,, \qquad &I_{\cal F}[{\cal F}_2] &=0 \,,
\end{align}
and therefore $\Gamma_{\rm cusp} (\lambda)= \lambda^2 + \mathcal{O}(\lambda^4)$, in agreement with \eqref{Gamma cusp conjecture ABJM-weak coupling-2} and the two-loop Feynman diagram computation of \cite{Griguolo:2012iq}.

Let us now discuss the computation of the four-loop contribution $\Gamma_{\rm cusp}^{(4)}$ from the $L=4$ integrated negative geometries computed in Section \ref{sec three loop results}. To do so it is useful to notice that, as in the ${\cal N}=4$ sYM case,
\begin{align}
\label{F functional-zp log a}
I_{\cal F}[z^p \log^a(z)] &= \lim_{\zeta \to 0} \frac{\partial^a}{\partial \zeta^a} \, I_{\cal F}[z^{p+\zeta}] \,.
\end{align}
In particular, this allows us to get
\begin{equation}
\label{Gamma H}
    I_{\cal F} \left[ z^{-1/4}\sqrt{1+z} \left( {\cal H}(z) + {\cal H} \left( \frac{1}{z} \right) \right) \right]= \frac{4\pi^2}{3} \,.
\end{equation}
Then, applying \eqref{F functional-zp} and \eqref{F functional-zp log a} to the different $L=4$ integrated negative geometry diagrams we arrive at
\begin{align}
I_{\cal F} \left[ {\cal F}_3^{\rm Ladd} \right]&= \frac{2\pi^2}{3} \,, \\
I_{\cal F} \left[ {\cal F}_3^{\rm Star} \right]&= \frac{\pi^2}{3} \,, \\
I_{\cal F} \left[ {\cal F}_3^{\rm Box} \right]&= 0 \,.
\end{align}
Interestingly, the box diagram has a vanishing contribution to the cusp anomalous dimension. It would be interesting to see if this result continues to be valid for higher-loop ``loops of loops'' diagrams. Combining all contributions we get
\begin{equation}
I_{\cal F} \left[ {\cal F}_3 \right]=-I_{\cal F} \left[{\cal F}_3^{\rm Ladd} \right]-I_{\cal F} \left[{\cal F}_3^{\rm Star} \right]+I_{\cal F} \left[{\cal F}_3^{\rm Box} \right] \,.
\end{equation}
Therefore, we obtain that the four-loop contribution to the cusp anomalous dimension of ABJM is
\begin{equation}
\label{four loop gamma cusp}
\Gamma_{\rm cusp}^{(4)}=-\pi^2 \,.
\end{equation}
This constitutes the first direct four-loop computation of the cusp anomalous dimension of the ABJM theory, and shows perfect agreement with the integrability-based proposal \eqref{Gamma cusp conjecture ABJM-weak coupling-2}.

Finally, let us end this section by discussing an integral representation of the $I_{\cal F}$ functional. To that end, it is useful to note that the result \eqref{F functional-zp} can be rewritten as
\begin{align}
\label{F functional integral -1}
I_{\cal F}[z^p] &= - \frac{4}{B\left( \frac{3}{4}+p, \frac{3}{4}-p \right)} \,,
\end{align}
where $B$ is the well-known Euler Beta function. Then, we can use the identity
\begin{equation}
\frac{1}{B(a,b)}= b \int_{c-i \infty}^{c+i \infty}  \frac{dz}{2\pi} \, z^{-a} (1-z)^{-1-b} \,, \qquad 0<c<1 \,, \quad  {\rm Re}(a+b)>0 \,,
\end{equation}
to get
\begin{align}
\label{F functional integral -2}
I_{\cal F}[z^p] &= -4 \left( \frac{3}{4}-p \right) \int_{c-i \infty}^{c+i \infty}  \frac{dz}{2\pi} \, z^{-3/4} (1-z)^{-7/4} \, \left[ \frac{1-z}{z} \right]^p \,.
\end{align}
Therefore,
\begin{align}
\label{F functional integral -3}
I_{\cal F}[{\cal F}] &= - \int_{c-i \infty}^{c+i \infty}  \frac{dz}{2\pi} \, z^{-3/4} (1-z)^{-7/4}\, \left[ 3 \, {\cal F}\left(\frac{1-z}{z}\right) + 4 \, z (1-z) \, \frac{d{\cal F}}{dz}\left(\frac{1-z}{z}\right)  \right] \,,
\end{align}
for $0<c<1$. One of the advantages of this representation of the $I_{\cal F}$ is that it is more amenable for numerical computations of the cusp anomalous dimension. As an example, we have tested \eqref{F functional integral -3} at $L=4$ loops, finding excellent numerical agreement with \eqref{four loop gamma cusp}.

\section{Conclusion}
\label{sec conclusion}

We have studied the IR-finite functions that arise when performing a three-loop integration over the four-loop integrand for the logarithm of scattering amplitude in the ABJM theory. With a focus on the four-particle case, we have performed the corresponding loop integrals by means of the differential equations method \cite{Henn:2013pwa}, therefore providing an application of this method to a three-dimensional theory. Along the lines of \cite{Henn:2023pkc,He:2023exb}, the $L$-loop integrated negative geometries splits into a parity-even term, proportional to a function ${\cal F}_{L-1}$, and a parity-odd contribution, described by a function ${\cal G}_{L-1}$. Extending the ideas of \cite{Henn:2023pkc,He:2023exb}, our results show that both the parity-even function ${\cal F}_{2k+1}$ and the parity-odd function ${\cal G}_{2k}$ have transcendentality degree $2k$ for $k\leq 1$. Moreover, we have found that the leading singularities are surprisingly simple in the limit in which the unintegrated variable goes to infinity. More precisely, in this frame the leading singularities are restricted to the set $\{ s\sqrt{t}, \, t\sqrt{s}, \, \sqrt{st(s+t)} \}$, up to the loop order we have studied. Furthermore, we have found that in the Euclidean region $z>0$ the signs of ${\cal F}_{2k+1}(z)$ and ${\cal G}_{2k}(z)$ alternate with $k$ for $k \leq 1$, suggesting a pattern for all $k$. This generalizes the results found for the ${\cal N}=4$ super Yang Mills theory in \cite{Arkani-Hamed:2021iya,Chicherin:2022zxo}, both at the four- and at the five-particle case. Finally, starting from the integrated negative geometries we have provided a direct computation of the four-loop contribution to the cusp anomalous dimension $\Gamma_{\rm cusp}$ of ABJM, in agreement with the integrability-based prediction that follows from \cite{Gromov:2008qe,Ahn:2008aa}.

Let us conclude by discussing some of the interesting questions that arise from our results. Firstly, it would be interesting to further extend the computation of the integrated negative geometries of ABJM to higher loops. While the integration of the $L=5$ order is expected to give a vanishing contribution to the cusp anomalous dimension, reaching the $L=6$ order would serve as the first six-loop test of the all-loop integrability-based conjecture for $\Gamma_{\rm cusp}$ \cite{Gromov:2008qe,Ahn:2008aa}. In turn, it would allow for a $\mathcal{O}(\lambda^5)$ test of the current all-loop proposal for the interpolating function $h(\lambda)$ of ABJM \cite{Gromov:2014eha}, that appears in all integrability-based results performed in this theory. It would be interesting to explore if a bootstrap analysis \cite{Dixon:2011pw,Dixon:2011nj,Dixon:2014voa,Dixon:2015iva,Caron-Huot:2016owq,Drummond:2014ffa,Dixon:2016nkn} could give access to these higher-loop integrated results. Another promising way to explore higher-loop integrated negative geometries relies in the Laplace-operator trick studied in \cite{Arkani-Hamed:2021iya}, which in the ${\cal N}=4$ sYM case provided an all-loop sum of ladder and tree diagrams.

Another question to address is the structure of higher-point negative geometries. To that end, the six- and eight-point integrands presented in \cite{He:2023rou} could serve as a starting point. It would be interesting to see if some of the properties found for the four-particle case also extend to the higher-multiplicity integrated results, such as the sign patterns and the apparent simplicity of the leading singularities. Furthermore, an interesting problem to study would be the description of the leading singularities of the $n$-point integrated negative geometries in terms of a Grassmannian formula \cite{Arkani-Hamed:2009ljj,Arkani-Hamed:2012zlh,Huang:2013owa,Huang:2014xza}, as done in \cite{Chicherin:2022bov} for an arbitrary number of particles and in the context of the ${\cal N}=4$ sYM theory.

Finally, one interesting direction to explore is the computation of the last loop integral that steps in the way between the integrated negative geometries and the logarithm of the corresponding scattering amplitude. Starting from our $L=4$ result, that integration would give the four-loop four-particle scattering amplitude of ABJM, which is currently unknown\footnote{Up to $L = 3$ loops, ABJM amplitudes have been studied in \cite{Gang:2010gy,Chen:2011vv, Bargheer:2010hn, Bargheer:2012cp, Bianchi:2012cq,Brandhuber:2012wy,Caron-Huot:2012sos,Bianchi:2013pfa,Bianchi:2014iia,He:2022lfz,Bianchi:2011dg,Bianchi:2013iha}.}.

\section*{Acknowledgements}

We would like to especially thank Johannes Henn for insightful discussions and guidance. Moreover, we thank Song He, Chia-Kai Kuo, Jungwon Lim, Antonela Matija\v{s}i\'{c}, Julian Miczajka and Giulio Salvatori for useful discussions. 
We would also like to thank Diego Correa for discussions and useful comments on the manuscript. ML is supported by fellowships from CONICET (Argentina) and DAAD (Germany).
This research received funding from the European Research Council (ERC) under the European Union's Horizon 2020 research and innovation programme (grant agreement No 725110), {\it Novel structures in scattering amplitudes}. 
\\[0.2cm]
\textbf{Note added}\\
While this work was in progress we were informed that similar questions were being addressed independently by Zhenjie Li \cite{LiToAppear}. We would like to thank him for correspondence and for confirming agreement with our $L=4$ results, see \eqref{F total} and \eqref{G total}.

\appendix

\section{Conformal invariance of integrated results}
\label{app conformal invariance}

In this section we will show that the integrated negative geometries (including both the leading singularities and the transcendental functions) are conformally invariant in the $x_5 \to \infty$ limit and after normalization with the three-dimensional Parke-Taylor factor (or any other factor of dimension $-2$ in energy units).

To start with, let us recall that in $d=3$ dimensions the spinor-helicity variables are defined as
\begin{equation}
\label{3d momentum matrix}
p^{ab}=\left( \begin{array}{cc}
p^0-p^1 & p^2 \\
p^2 & p^0+p^1
\end{array}
\right)= \lambda^a \lambda^b \,.
\end{equation}
In this context, Lorentz invariants can be constructed as
\begin{equation}
    \langle ij \rangle= \epsilon_{ab} \lambda_i^a\lambda_j^b \,,
\end{equation}
where we are using $\epsilon_{12}=1$, and therefore
\begin{equation}
s=(p_1+p_2)^2=-\langle 12 \rangle^2 \,, \qquad t=(p_2+p_3)^2=-\langle 23 \rangle^2 \,.
\end{equation}
Moreover, the generator of special conformal transformations is given as \cite{Bargheer:2010hn}
\begin{equation}
    K_{ab}=\sum_{i=1}^4 \partial_a^i \partial_b^i \;.
\end{equation}
Let us consider a general function $f(\alpha,\beta)$, where $\alpha= \langle 12 \rangle$ and $\beta= \langle 23 \rangle$. Then,
\begin{equation}
\label{symmetry eqs 1}
K_{ab} f= \partial_{\alpha}^2 f \left( \langle 1|_a \langle 1|_b + \langle 2|_a \langle 2|_b \right) + \partial_{\beta}^2 f \left( \langle 2|_a \langle 2|_b + \langle 3|_a \langle 3|_b \right)- \partial_{\alpha}\partial_{\beta} f \left( \langle 1|_a \langle 3|_b + \langle 1|_b \langle 3|_a \right) \,,
\end{equation}
where we are defining $\langle i|_a :=\epsilon_{ab}\lambda_i^b$. It is useful to note that to completely characterize the $K_{ab} f$ matrix we only need to compute the matrix elements $K_{ab} f \, |1 \rangle^a |1 \rangle^b$, $K_{ab} f \, |2 \rangle^a |2 \rangle^a$ and $K_{ab} f \, |1 \rangle^a |2 \rangle^b$, where $|i \rangle^a :=\lambda_i^a$. Consequently, \eqref{symmetry eqs 1} is equivalent to 
\begin{equation}
\label{conf inv eqs}
   \left\{ \begin{array}{l}
         \alpha^2 \partial^2_{\alpha} f-\beta^2 \partial^2_{\beta} f=0 \,,  \\
         \alpha^2 \partial^2_{\alpha} f+\beta^2 \partial^2_{\beta} f+ 2 \alpha \beta \partial_{\alpha} \partial_{\beta} f=0 \,, \\
         \beta \partial^2_{\beta}f +\alpha \partial_{\alpha} \partial_{\beta} f =0\,.
    \end{array}
    \right.
\end{equation}
At this point we should note that by dimensional analysis we can always write
\begin{equation}
\label{normalized neg geom}
{\cal V}(\alpha, \beta) \, \left( \prod_{j=6}^{4+L} \int \frac{d^3 X_j}{i\pi^{3/2}}  \, {\cal L}_L \right) \bigg|_{x_5 \to \infty}= \alpha \, {\cal N} \left( \frac{\beta}{\alpha}\right) \,,
\end{equation}
for some function ${\cal N}$ and where ${\cal V}(\alpha, \beta)$ is some factor with dimension $-2$ in units of energy (for example, the three-dimensional Parke-Taylor factor $ {\rm PT}^{(3)}= \frac{1}{\alpha \beta} $). We see that \eqref{normalized neg geom} will always satisfy the constraints \eqref{conf inv eqs}, regardless of the function ${\cal N}$. Consequently, we conclude that the integrated negative geometries of ABJM are conformally invariant in the $x_5 \to \infty$ limit and after normalization with a factor of dimension $-2$ in units of energy. This generalizes the result presented in \cite{Henn:2023pkc}, where it was found that the leading singularities of the integrated negative geometries of ABJM were conformally invariant in the $x_5 \to \infty$ frame and after normalization with the three-dimensional Parke-Taylor factor.

\bibliographystyle{JHEP}
\bibliography{ABJMneggeom.bib}

\end{document}